\documentclass[12pt]{article}
\usepackage{graphicx}

\def\MPIK{Max-Planck-Institut f\"ur Kernphysik\\ D-69117 Heidelberg, GERMANY}

\def\Title#1{\begin{center} {\Large #1 } \end{center}}
\def\Author#1{\begin{center}{ \sc #1} \end{center}}
\def\Address#1{\begin{center}{ \it #1} \end{center}}

\newenvironment{Abstract}{\begin{quotation}  }{\end{quotation}}
\newenvironment{Presented}{\begin{quotation} \begin{center} 
             PRESENTED AT\end{center}\bigskip 
      \begin{center}\begin{large}}{\end{large}\end{center} \end{quotation}}

\begin{document}
\begin{titlepage}

\vfill
\Title{Sterile Neutrinos: Reactor Experiments}
\vfill
\Author{Christian Buck}
\Address{\MPIK}
\vfill
\begin{Abstract}
Nuclear reactors are strong, pure and well localized sources of electron antineutrinos with energies in the few MeV range. Therefore they provide a suitable environment to study neutrino properties, in particular neutrino oscillation parameters. Recent predictions of the expected antineutrino flux at nuclear reactors are about 6\% higher than the average rate measured in different experiments. This discrepancy, known as the reactor antineutrino anomaly, is significant at the 2.5$\sigma$ level.   

Several new experiments are searching for the origin of this observed neutrino deficit. One hypothesis to be tested is an oscillation to another neutrino state. In a three flavor model reactor neutrinos do not oscillate at baselines below 100~m. Hence, if such an oscillation is observed, it would imply the existence of at least one light sterile neutrino state not participating in weak interactions. Such a discovery would open the gate for new physics beyond the Standard Model.    

\end{Abstract}
\vfill
\begin{Presented}
NuPhys2016, Prospects in Neutrino Physics\\
Barbican Centre, London, UK,  December 12--14, 2016
\end{Presented}
\vfill
\end{titlepage}

\section{Introduction}
Many neutrino experiments in the last two decades have shown that neutrinos oscillate. This quantum mechanical effect is driven by the existence of different neutrino mass eigenstates which are not identical to the three known flavor eigenstates. Reactor neutrino experiments played a crucial role in measuring the oscillation parameters. The KamLAND experiment in Japan improved our knowledge on the ``solar'' mixing parameters~\cite{Eguchi:2002dm}. The smallest of the three neutrino mixing angles, $\theta_{13}$, was recently determined by the $\sim$1~km baseline experiments Double Chooz~\cite{Abe:2011fz}, Daya Bay~\cite{An:2012eh} and RENO~\cite{Ahn:2012nd}. 

Nuclear reactors produce electron antineutrinos in the $\beta$-decay of the neutron rich fission fragments in the reactor core. On average, there are six antineutrinos emitted per fission, resulting in an isotropic flux of more than $10^{20}$neutrinos per GW of thermal power in every second. Besides a precise knowledge of the thermal power of the reactor and the distance to the detector, the time dependent fractional fission rates of the four main isotopes $^{235}$U, $^{238}$U, $^{239}$Pu and $^{241}$Pu need to be modelled. However, the main uncertainty in the flux predictions is coming from the neutrino energy spectrum of each fission isotope.     

The standard reaction used to detect the electron antineutrinos is the inverse beta decay (IBD) on protons, typically in organic liquid scintillator (LS) detectors. In this reaction a coincidence signal of a prompt positron and a delayed neutron event is produced. The energy threshold for the IBD is at 1.8~MeV. The prompt positron deposits its kinetic energy in the neutrino detector and finally annihilates with an electron which produces two 511~keV gammas. The ``visible energy'' in the detector is about 0.8~MeV less than the energy of the incoming neutrino. The neutron produced in the IBD reaction is mainly captured on hydrogen for the case of an unloaded LS. In this delayed event a 2.2~MeV gamma is emitted. For better background discrimination the LS is often doped with gadolinium (Gd) increasing the gamma energy upon neutron capture to about 8~MeV and reducing the coincidence time significantly. For improved event localization the Gd can be replaced by $^6$Li, which decays into a triton and an alpha particle after neutron capture.     

\section{The reactor antineutrino anomaly}
In the context of the $\theta_{13}$ experiments, the neutrino flux predictions at nuclear reactors were re-evaluated~\cite{Mueller:2011nm, Huber:2011wv}. These new calculations revealed an increase of the flux prediction of few percent as compared to the Schreckenbach et al.~predictions~\cite{Schreckenbach:1985ep, Hahn:1989zr}, which provided the reference spectra before. All different conversion methods for the $^{235}$U, $^{239}$Pu and $^{241}$Pu neutrino spectra rely on the measured beta spectra at the Institut Laue-Langevin (ILL) in the early 1980ies~\cite{VonFeilitzsch:1982jw}. The normalization shift of the recent flux predicitions is the main contribution to the observed discrepancy to the reactor neutrino data. However, there are other effects which enhanced the significance of the data to prediction difference. Among those are the addition of non-equilibrium effects in the calculations and a shift of the measured neutron lifetime within the last 30 years. The neutron lifetime has a direct impact on the calculations of the IBD cross-section.    

This reactor antineutrino anomaly~\cite{Mention:2011rk} of an electron antineutrino deficit observed few meters from the reactor core, could be interpreted as an oscillation into another neutrino state. However, this would require a splitting between the squared neutrino mass eigenstates involved in this oscillation which is at the order of 1~eV$^2$. This is significantly higher than the known mass splittings in the three flavor paradigm and therefore requires at least a fourth neutrino state. Since there can only be three active light neutrinos participating in weak interactions, a fourth generation would have to be sterile. The allowed oscillation parameter region that could explain the reactor anomaly is similar to the one of the longstanding LSND anomaly~\cite{Aguilar:2001ty}. Moreover, such an oscillation could also explain the slightly lower than expected values found in the neutrino source calibration runs of the solar neutrino experiments GALLEX~\cite{Kaether:2010ag} and SAGE~\cite{Abdurashitov:2005tb}. There are many reactor neutrino experiments, which just started or are currently under construction, addressing the question of the light sterile neutrinos. Some of them will be discussed in more detail in the next section. 

The recent $\theta_{13}$ experiments also observed spectral distortions in the neutrino spectrum known as reactor ``shape anomaly''~\cite{Abe:2014bwa, An:2015nua, RENO:2015ksa}. The main feature of this distortions is an excess in the measured reactor neutrino spectrum as compared to the predicted shape around 5~MeV. Sterile neutrinos would not explain this shape anomaly, which is more likely attributed to nuclear and reactor physics. The new generation of experiments might be able to identify if this shape distortion is common to all fission isotopes or caused by only part of them. Some experiments are operated at highly enriched reactors (HEU) in which mainly $^{235}$U fissions contribute. Other experiments use commercial reactors in which several U and Pu isotopes contribute (LEU reactors). If the neutrino spectra of HEU and LEU reactors are compared it might show if the 5~MeV excess is solely due to the $^{235}$U contributions or similar for all isotopes.  

\section{Very short baseline experiments} 
Experiments search for sterile neutrinos at nuclear reactors by checking for oscillation effects in the energy spectrum, in the neutrino rate at different baselines or both. The detector requirements are low background environment, high energy resolution, precise energy scale knowledge and a baseline in the 10~m range or even below. Segmentation and modularity also help to improve the sensitivity. On the reactor side a strong, but compact core is desirable. 

\subsection{Nucifer}  
The Nucifer experiment~\cite{Boireau:2015dda} was running at a baseline of about 7~m from the 70~MW OSIRIS HEU reactor at CEA Saclay in France. Nucifer was designed and motivated in the context of reactor monitoring and safeguard applications. There is interest from the International Atomic Energy Agency (IAEA) in a continuous monitoring of the fissile content in a nuclear reactor through non-intrusive techniques to reduce the risk of proliferation of nuclear weapons. With Nucifer it was demonstrated that the presence of 1.5~kg of Pu could be detected inside a Osiris-like core with a $\sim$1~ton detector opening the possibility of a first societal application of neutrino physics. 

In principle, Nucifer can also be used to probe the reactor antineutrino anomaly. However the lack of precision in the neutrino flux prediction and the high rate of accidental background do not allow for strong statements regarding the sterile neutrino search. Nevertheless, a new data point provided by Nucifer is now included in global analyses giving further constraints on the allowed parameter space.   

\subsection{NEOS}
As the Nucifer detector also the NEOS detector~\cite{Ko:2016owz} has an unsegmented target volume consisting of about 1~m$^3$ of Gd-loaded organic liquid scintillator (Gd-LS). The main differences between Nucifer and NEOS are the reactor and the signal to background ratio. NEOS is operated at a commercial LEU reactor with a thermal power of 2.8~GW providing a strong neutrino flux. The signal to background ratio at the site of the detector at a baseline of 25~m is above 20. NEOS reported an impressive precision on their energy scale with a systematic uncertainty of only 0.5\%.  

Due to the high statistics of about 2000 detected antineutrinos per day it was possible with NEOS to confirm the spectral distortion around 5~MeV with high significance. The collaboration reported that their data could exclude the parameter space of sin$^2(2\theta_{14})$ below 0.1 for $\Delta$m$^2_{41}$ ranging from 0.2~eV$^2$ to 2.3~eV$^2$ at a confidence level of 90\%~\cite{Ko:2016owz}. This exclusion area already disfavours some of the best fit points in global analyses. The minimum $\chi^2$ value in a 3+1 hypothesis was found for the pair (sin$^2(2\theta_{14})$, $\Delta$m$^2_{41})= (0.05, 1.73$~eV$^2$), but overall no strong evidence for 3+1 neutrino oscillations were observed in this experiment.

\subsection{DANSS}
In the DANSS experiment \cite{Alekseev:2016llm} a highly segmented detector with 1~m$^3$ of plastic scintillator is operated at the 3~GW reactor of the Kalinin nuclear power plant in Russia. The high power of this commercial reactor in combination with the rather short baseline of $10-12$~m provides a high statistics neutrino signal of about 5000 IBD events per day. The detector setup can be moved up and down to test different neutrino oscillation lengths at 3 different positions. The site has a 50~m.w.e.~shielding which limits the cosmic background to 5\% of the neutrino signal. The basic element of the DANSS detector are plastic scintillator strips (1x4x100~cm$^3$) co-extruded with a white layer for light containment. This polystyrene based coating contains 6\% of Gd oxide corresponding to 0.35~wt.\% of pure Gd with respect to the target material.  

The concept was tested with DANSSino, a prototype detector with 2 modules. The elements are designed identically as the 50 modules used in the full scale detector. Each module consists of 50 scintillator strips. With DANSSino it was possible to study the background level and even reactor antineutrinos could be measured at a rate of 70 IBDs per day with a signal to noise ratio around unity. The full scale experiment started data taking in 2016. 

\subsection{Neutrino-4}
Another experiment in Russia is the Neutrino-4 project~\cite{Serebrov:2016wzv} at the 100~MW SM-3 reactor. As in Nucifer or NEOS the IBD reactions are detected in a Gd-LS target with a total volume of 3~m$^3$. The detector is moveable with the closest position to the reactor at about 7~m and the longest baseline at 11~m. The expected neutrino rate is 1000~events per day. The main challenge in Neutrino-4 is the control of the background level. In a 350~liters prototype the rate due to cosmic background was about 4 times higher than the neutrino signal. Now the project is running with the full scale detector as well. The detector is surrounded by layers of active (12~cm scintillator plates) and passive (60~tons of steel, lead and borated polyethylene) shielding.   

\subsection{Stereo}
The Stereo experiment at the Institut Laue-Langevin (ILL) in Grenoble, France, is measuring neutrinos 10~m away from a compact 58~MW research reactor highly enriched in $^{235}$U. The 2.2~m long detector has a neutrino target segmented in six identical cells, all of them filled with a Gd-LS. Stereo aims to measure the relative distortions of the neutrino energy spectrum in these cells caused by neutrino oscillation at different distances from the reactor core. The more than 1800 liters of neutrino target are surrounded by another 2100 liters of Gd-free LS to detect escaping gammas. The produced scintillation light is collected by a set of 48 photomultiplier tubes (PMTs) of 8~inch diameter which are separated from the LS by an acrylic buffer and n-dodecane. 

As a consequence of the reactor vicinity and the surrounding neutron beam lines, the STEREO environment has a rather high background level of neutrons and gammas. For that reason a heavy shielding made of B$_4$C, lead and borated polyethylen surrounds the detector. In addition, a water-Cherenkov veto on top of the detector tags cosmic muons at the shallow depth of the site. The STEREO experiment started data taking in November 2016 and detects 400~neutrinos per day in reactor-on phases. After a period of 2~years Stereo should be able to probe the main part of the allowed parameter region of the reactor antineutrino anomaly.

\subsection{SoLid}
The sterile neutrino search in SoLid \cite{Ryder:2015sma} will be performed between 5.5 and 10~m from the highly enriched uranium core of the BR2 reactor in Belgium. The experimental concept is based on precise localisation of the IBD events combined with a high neutron-gamma discrimination capability. To achieve this goal, a composite scintillator design is applied. The neutrino target is made out of 5~cm cubes of polyvinyl toluene scintillator (PVT). The energy depositions of annihilation gammas  in neighbouring cubes can be used to tag the prompt positron of the IBD interaction. On one face of each PVT cube there is a neutron sensitive layer of $^6$LiF:ZnS(Ag). The cubes are optically separated by reflective Tyvek for optimized light collection. After thermalization in the PVT cube the IBD neutron has 50\% probability to be captured on a $^6$Li nucleus. Such an interaction results in the production of an alpha and a triton particle. Most of the 4.78~MeV energy of those two particles is deposited inside the ZnS(Ag) inorganic scintillator. The time profile of the photon production in the ZnS is much slower ($\mu$s scale) than the one of PVT signals (ns scale). These characteristic time signatures can be used for further background discrimination. The technology was successfully tested in a 288~kg prototype module. The PVT cubes were arranged in 9 frames each with 16x16 cubes.        

\subsection{Prospect}
In the phase I of the PROSPECT experiment~\cite{Ashenfelter:2015uxt} short baseline oscillations are studied in a movable detector at baselines from $7-12$~m. The 85~MW High Flux Isotope Reactor (HFIR) at the Oak Ridge National Laboratory in the US provides a pure $^{235}$U neutrino spectrum. This spectrum could be measured with a precision in a segmented LS detector loaded with $^6$Li. In the detector design the target volume is 3000~l with 120 optically separated moduls (each 15x15x120~cm$^3$) including double ended PMT readout. As in the other similar experiments the key challenge of Prospect is background reduction. Efficient background suppression should be achieved using vertex information and pulse shape discrimination separating electron/gamma like events from heavy recoils. In this way, a signal to background ratio of better than 1:1 is predicted.

Prospect was undergoing a staged approach in prototype and shielding developments. Tests on the Li-loaded LS were started on the 100~ml scale to characterize the liquid properties. First background studies were then performed in a 1.7~l cell which was upscaled to 23~l for further LS and background studies. In Spring 2016 a prototype with 2 moduls with the dimensions of the full scale detector was developed. The measurements with the 3~ton detector is expected to commence in 2017 and should run for 3~years. Afterwards in a phase II of the experiment PROSPECT could be upgraded with a second detector at $15-19$~m baseline for further improvement of the sensitivity.  

\section{Summary}
The tension between the observed rate of reactor antineutrinos at short baselines with most recent flux predictions triggered the development of several new experiments searching for oscillations into sterile neutrinos. To obtain a convincing sterile neutrino signature the main features of such experiments should include short baselines ($\sim$10~m), segmentation and/or movability as well as effective background suppression techniques.     

\begin{table}[t]
\begin{center}
\begin{tabular}{l|cccccc}  
Experiment &  P$_{th}$ [MW] & L [m] & depth [mwe] & m [t] & technique & S/B \\ \hline
Nucifer & 70 & 7 & 12 & 0.8 & Gd-LS & $<$1\\
Neos & 2700 & 25 & 20 & 1 & Gd-LS & 22\\ 
DANSS & 3000 & 9--12 & 50 & 0.9 & Gd-PS & $\sim$20\\
Neutrino-4 & 100 & 6--11 & 5--10 & 1.5 & Gd-LS & $<$1\\
Stereo & 57 & 9--11 & 10 & 1.7 & Gd-LS & $>$1\\
Solid & 100 & 6--11 & 10 & 1.6 & $^6$Li-PS & $\sim$3\\ 
Prospect & 85 & 7--12 & $\sim$5 & 3 & $^6$Li-LS & $>$1\\ \hline
\end{tabular}
\caption{Comparison of running and upcoming sterile neutrino experiments listing the thermal power of the reactor (P$_{th}$), baseline (L), shielding, target mass (m), detection technique and signal to background ratio (S/B). All of the experiments in the table except Nucifer and Neos are using segmented detectors.}
\label{VSBE}
\end{center}
\end{table}

In Table \ref{VSBE} several experiments are listed which either published results recently, started data taking or are currently under construction.  With those strong efforts in the field the next years should allow to prove the existence of light sterile neutrinos or exclude the currently allowed region for $\sim$eV mass splittings. Since several of these reactor experiments are operated at compact cores highly enriched in $^{235}$U, high precision measurements of the associated neutrino spectrum will be obtained. This should give new insights into the observed distortions in the neutrino spectra of the km baseline $\theta_{13}$ experiments as compared to the predictions.

\end{document}